\documentclass[12pt]{article}
\usepackage{amsmath,amssymb,amsfonts,epsfig,graphicx,euscript}%
\newcommand{\bse}{\begin{subequations}}
\newcommand{\ese}{\end{subequations}}
\newcommand{\be}{\begin{equation}}
\newcommand{\ee}{\end{equation}}
\newcommand{\bea}{\begin{eqnarray}}
\newcommand{\eea}{\end{eqnarray}}
\newcommand{\ba}{\begin{array}}
\newcommand{\ea}{\end{array}}

\begin{document}
\hfill%
\vbox{
    \halign{#\hfil        \cr
           IPM/P-2011/004\cr
                     }
      }
\vspace{1cm}
\begin{center}
{ \Large{\textbf{Non-commutative holographic QCD}\\
\textbf{and}\\
\textbf{DC conductivity}\\}} \vspace*{2cm}
\begin{center}
{\bf M. Ali-Akbari}\\%
\vspace*{0.4cm}
{\it {School of Physics, Institute for Research in Fundamental Sciences (IPM)\\
P.O.Box 19395-5531, Tehran, Iran}}  \\
{E-mails: {\tt aliakbari@theory.ipm.ac.ir}}\\%
\vspace*{1.5cm}
\end{center}
\end{center}

\vspace{.5cm}
\bigskip
\begin{center}
\textbf{Abstract}
\end{center}
In this paper we consider non-commutative Sakai-Sugimoto model
\cite{Sakai} by using D8-branes probing the background generated by
D4-branes with an NSNS B-field turned on. We study response of the
system to external electric field which could be parallel to
orthogonal to the background B-field. We compute the conductivity as
a function of temperature and non-commutativity parameters.
Non-commutativity effect, depending on the relative orientation of
external and background electromagnetic fields, may increase or
decrease the conductivity.
\newpage

\section{Introduction}
According to the Anti-de Sitter/conformal field theory (AdS/CFT)
conjecture, IIB string theory on $AdS_5\times S^5$ background
is dual to $D=4\ {\cal{N}}=4\ SU(N_c)$ super Yang-Mills theory (SYM)
\cite{Maldacena:1997re}. In the large $N_c$ and large 't
Hooft coupling $\lambda=g^2_{YM}N_c$ limit, the SYM theory is dual to IIb
supergravity which is low energy effective theory of superstring
theory. As a result a strongly coupled thermal SYM theory corresponds to supergravity
in an AdS black brane background where the SYM theory temperature is
identified with the Hawking temperature of the AdS black hole
\cite{Witten:1998zw}.

AdS/CFT idea has been applied to study different aspects of strongly
coupled gauge theories. Recently the application of this duality in
condensed matter physics (called AdS/CMT) has been studied
\cite{Review} . This duality is very useful to study certain
strongly coupled systems in CMT by holographic AdS/CFT techniques
and to understand better their properties. Of interest is to compute
response of the system to the external forces or fields, in
particular the response to the external electric field, the
conductivity \cite{Karch}.

There are two important holographic models describing quantum
chromodynamics (QCD) which are based on dynamics of D3-D7 or D4-D8
(the Sakai-Sogimoto model) \cite{Sakai,Aharony} systems. Various
different aspects of these models such as phase diagram, chiral
symmetry breaking, hadronic spectrum and response to the external
electric-magnetic fields have been extensively studied in these
contexts in the literature
\cite{Karch,AliAkbari,OBannon,Bergman,Erdmenger}. In order to
compute conductivity in D3-D7 system, on the gravity side, $N_f$
flavor D7-branes are introduced in the probe limit, \textit{i.e.}
$N_f\ll N_c$ and hence AdS background is left unchanged. On the
gauge theory side introduction of $N_f$ D7-branes amounts to
introducing ${\cal{N}}$=2 hypermultiplets in the fundamental
representation of the gauge group. These hypermultiplets which are
in $U(N_f)$ representation may be associated with the open strings
between the D7 flavor branes and the D3-branes of the background.
The local $U(N_f)$ symmetry on the D7-branes corresponds to global
$U(N_f)$ symmetry whose $U(1)_B$ subgroup may be identified with
baryon number. Non-dynamical electric and magnetic fields can be
coupled to $U(1)_B$ charge and we then expect a constant, nonzero
current.
Hence the conductivity tensor $\sigma_{ij}$ is identified by %
\be %
 <J_i>=\sigma_{ij}E_j\,.
\ee %
Electric and magnetic fields produce diagonal and off-diagonal
elements in conductivity tensor respectively. In fact on the gravity
side currents, electric and magnetic fields are introduced as
nontrivial gauge fields living on the D7-branes \cite{OBannon}.

In D4-D8 system, D8-branes have the same role as D7-branes in
pervious case and hence open strings stretched between the $N_f$
flavor D8-branes and $N_c$ color D4-branes are associated with
fields in fundamental representation in the CFT side. However D3-D7
system is not a chiral model, D4-D8 model enjoys chiral symmetry and
explains chiral symmetry breaking \cite{Sakai,Aharony}. Response of
the system to the external electromagnetic field and conductivity in
D4-D8 system were also discussed in \cite{Bergman}.

Non-commutative gauge theories naturally appear on the D-branes with
a background  NSNS B-field on them. Explicitly consider a system of
D$p$-branes with a constant NSNS B-field along their worldvolume
directions. By taking a low energy limit , closed strings decouple
and the resulting action for open strings is the non-commutative
gauge theory \cite{Ardalan}. It is possible to extend the AdS/CFT
dictionary to the cases involving background B-field in the gravity
side and non-commutative gauge theory in the CFT side
\cite{Hashimoto:1999ut,Maldacena:1999mh}.

What we will consider in this paper is the effect of the NSNS
B-field on the conductivity in the Sakai-Sugimoto model which will
be reviewed in section 2. In section 3, a number of D8-branes are
embedded in the "non-commutative geometry" corresponding to the
non-commutative QCD, hence building a non-commutative Sakai-Sugimoto
model. We then find conductivity in non-commutative gauge theories
at low and high temperatures in this non-commutative Sakai-Sugimoto
model. In next section we consider a more general non-commutative
Sakai-Sugimoto background in which two independent components of NS
B-field ($B_{01},B_{23}$) are turned on and study the conductivity.
The last section is devoted to conclusion.

\section{A brief review on Sakai-Sugimoto model}
In this section, we review holographic QCD background
(Sakai-Sugimoto model) at low and high temperature
\cite{Sakai,Aharony}. The holographic QCD model is constructed from
the near horizon limit of a set of $N_c$ D4-branes compactified on
a circle with an anti-periodic boundary condition for the adjoint
fermions. This makes the adjoint fermions massive and breaks
supersymmetry. Fermions in (anti-)fundamental representation are
introduced by $N_f$ (anti-)D8-branes intersecting the D4-brane at a
3+1 -dimensional defect. There is thus a global $U(N_f)\times
U(N_f)$ chiral symmetry from the worldvolume of D4-brane point of
view. We work in the prob limit, namely $N_f\ll N_c$, where flavour
branes do not backreact on the background.

At low temperature, the near horizon of D4-branes reads \cite{Aharony} %
\be\begin{split}\label{metric1} %
 ds^2&=(\frac{u}{R})^{3/2}\bigg(dt_E^2+dx^idx_i
 +f(u)dx_4^2\bigg)+(\frac{R}{u})^{3/2}\bigg(\frac{du^2}{f(u)}+u^2d\Omega_4^2\bigg),
\cr
 e^{\phi}&=g_s(\frac{u}{R})^{3/4},\ \ \ \ \ f(u)=1-\frac{u_k^3}{u^3},\ \ \ \ \ F_4=dC_3=\frac{2\pi N_c}{V_4}\epsilon_4,
\end{split}\ee %
where $t_E\ ({\rm{Euclidean\  time}}),\ x^i(i=1,2,3)$ and $x_4$ are
the directions along which the D4-branes are extended. $d\Omega_4^2,
\epsilon_4$ and $V_4=8\pi^2/3$ are the line element, the volume form
and the volume of a unit $S^4$, respectively. $R$ and $u_k$ are
constant parameters. $R$ is related to the string coupling $g_s$ and
string length $l_s$ as $R^3=\pi g_sN_cl_s^3$. The coordinate $u$ is
bounded from below by the condition $u\geq u_k$. In order to avoid a
singularity at $u=u_k$, $x_4$ must be a periodic
variable with period ${\cal{R}}$%
\be %
 2\pi{\cal{R}}=\frac{4\pi}{3}(\frac{R^{3}}{u_k})^{1/2}
\ee %

The high temperature background is given by \cite{Aharony} %
\be\begin{split}\label{metric2} %
 ds^2&=(\frac{u}{R})^{3/2}\bigg(f(u)dt_E^2+dx^idx_i
 +dx_4^2\bigg)+(\frac{R}{u})^{3/2}\bigg(\frac{du^2}{f(u)}+u^2d\Omega_4^2\bigg),\cr
 f(u)&=1-\frac{u_T^3}{u^3},\ \ \ \ \ \delta
 t_E=\frac{4\pi}{3}(\frac{R^{3}}{u_T})^{1/2}=\frac{1}{T}=\beta,\ \ \
 \ \ u_T=(\frac{4\pi T}{3})^2,
\end{split}\ee %
where $T$ is the background temperature and all the other parameters
are defined as before.

It was shown in \cite{Aharony} that this theory undergoes a
confinement-deconfinement phase transition at a temperature
$T_d=1/2\pi {\cal{R}}$. For quark separation obeying
$R>0.97{\cal{R}}$, the chiral symmetry is restored at this
temperature but for $R<0.97{\cal{R}}$ there is an intermediate phase
which is deconfined with broken chiral symmetry and the chiral
symmetry is restored at $T_{\chi SB}=0.154R$.
%The chiral symmetry
%breaking in non-commutative background was discussed in
%\cite{Nakajima}.

\section{Non-commutative conductivity}
The response of the D3-D7 system (dual to ${\cal{N}}=2$ SYM with $N_f$ flavor
hypermultiplets) to external electric field $E$ was originally discussed in \cite{Karch},
where it was found that in the massless limit the conductivity $\sigma$ is given by %
\be\label{Oconductivity} %
 \sigma^2=\frac{N_f^2N_c^2T^2}{16\pi^2}\sqrt{e^2+1}+\frac{d^2}{e^2+1}
\ee %
where $N_c$ and $T$ are related to radius and temperature of AdS black hole.
Moreover $e$ and $d$ are defined as %
\be %
  e=\frac{E}{\frac{\pi}{2}\sqrt{\lambda} T^2},\ \ \ \ \ d=\frac{D}{\frac{\pi}{2}\sqrt{\lambda} T^2}
\ee %
where $\lambda=g_s^2N_c$. $E$ and $D$ are external electric field
and (free) charge density introduced by non-trivial gauge field
living on D7-branes. One can eliminate the effect of free charge in
\eqref{Oconductivity} by setting $A_0$ to zero on branes
\cite{Karch}. Hence the first term in
\eqref{Oconductivity} in two different limits becomes %
\begin{itemize}
  \item Zero mass, zero density and zero external field
  \be\label{limit1} %
  \sigma=\frac{N_fN_c}{4\pi}T
  \ee %
  The factor $N_fN_c$ is number of charge carriers.
  \item Zero mass, zero density and zero temperature \footnote{$\frac{E}{\sqrt{\lambda}}$ is
  the natural combination which appears in the strong coupled SYM regime \textit{e.g.} \cite{Maldacena2}.}
    \be\label{limit2} %
  \sigma^2=\frac{N_f^2N_c^2}{8\pi^3}\frac{E}{\sqrt{\lambda}}
  \ee %
\end{itemize}

What we are going to do in the following subsections is to find
conductivity in low and high temperature within the non-commutative
Sakai-Sugimoto model. To do so, we add a number $N_f$ of D8-branes
filling all directions except $x_4$ in non-commutative background.
For simplicity $A_0$ is set to zero and therefore the effect of free
charge density does not appear in our result.

\subsection{Low temperature case}
The low temperature non-commutative background is given by
\footnote{Hereafter we consider non-commutative metrics in
units where the background $AdS$ radius is one.}
\cite{Nakajima} %
\be\begin{split}\label{Nmetric1} %
 ds^2&=
 u^{3/2}\bigg(h(u)(dt_E^2+dx_1^2)+dx_2^2+dx_3^2+f(u)dx_4^2\bigg)\cr
 &+u^{-3/2}\bigg(\frac{du^2}{f(u)}+u^2d\Omega_4^2\bigg),\cr
 h(u)&=\frac{1}{1+\theta^3u^3},\ \ \ \ \ e^{\phi}=g_su^{3/4}\sqrt{h(u)},\ \ \ \ \ B=B_{t1}=\theta^{3/2}u^3h(u),
\end{split}\ee %
where $\theta$ is non-commutative parameter. This metric reduces to
\eqref{metric1} by setting $\theta$ to zero.

The low energy effective action for $N_f$ D8-branes in an arbitrary
background is given by Dirac-Born-infeld (DBI) action \footnote{Note
also that Chern-Simon action does not contribute in \eqref{action} for our case.}%
\be\label{action} %
 S_{D8}=-T_8N_f\int d^9\xi e^{-\phi}\sqrt{\det\Big(g_{\mu\nu}+B_{\mu\nu}+(2\pi\alpha')F_{\mu\nu}\Big)}
\ee %
where $\mu$ is worldvolume index running from 0,...,9. $g_{\mu\nu}$ and $B_{\mu\nu}$ are induced metric and
induced Kalb-Ramond fields defined by %
\be\begin{split} %
 g_{\mu\nu}=G_{MN}\partial_\mu X^M\partial_\nu X^N, \cr
 B_{\mu\nu}=B_{MN}\partial_\mu X^M\partial_\nu X^N.
\end{split}\ee %
In our case $G_{MN}$ is non-commutative metric \eqref{Nmetric1} where $M,N=0,...,9$. $F_{\mu\nu}$
is field strength of the gauge field living on the brane. The brane tension is $T_8=\frac{2\pi}{(2\pi l_s)^9g_s}$
and $\phi$ is the dilaton field. Note that in this background Klab-Ramond  field is no longer
zero and it therefore contributes to DBI action.

Let us start with following ansatz for $A_1(u,t_E)$ in Euclidean space %
\be %
 A_1(t_E,u)=iEt_E+a_1(u).
\ee %
In static gauge, the DBI action for $N_f$ D8-branes in the low temperature background \eqref{Nmetric1} becomes %
\be\label{dbiaction}\begin{split} %
 S_{D8}=-\frac{8\pi^2}{3}N_fT_8\int dudt_E\frac{1}{\sqrt{h}}
 \sqrt{\hat{g}_{22}g_{33}\big[g_{tt}(\hat{a}_1'^2+g_{11}g_{uu})-g_{uu}(B^2+\hat{E}^2)\big]}\ ,
\end{split}\ee %
where $\hat{E}=(2\pi\alpha')E$, $\hat{g}_{22}=u^{1/2}g_{22}$ and $\hat{a}_1=(2\pi\alpha')a_1$. Since the above action will only depend upon $u$-derivative of
$a_1(u)$, we will have a conserved charge associated with $a_1(u)$ which is given by %
\be\label{nontermalcharge} %
 I_1\equiv\frac{1}{\sqrt{h}}\frac{{\cal{N}}(2\pi\alpha')^2\hat{g}_{22}g_{tt}g_{33}a_1^{\prime}}
 {\sqrt{\hat{g}_{22}g_{33}\big[g_{tt}(\hat{a}_1'^2+g_{11}g_{uu})-g_{uu}(B^2+\hat{E}^2)\big]}}.
\ee %
where \footnote{Note that by choosing $R=1$, we have $\lambda=4\pi
g_sN_c\alpha'^{1/2}=\frac{4}{\alpha'}.$}
\be %
 {\cal{N}}=\frac{8\pi^2}{3}N_fT_8=\frac{1}{6(2\pi)^6}N_cN_f\lambda^2.
\ee %
Solving \eqref{nontermalcharge} for the gauge field leads to%
\be\label{A1} %
 \hat{a}_1^{\prime}=\pm\sqrt{\frac{hg_{uu}\big(B^2+\hat{E}^2-g_{11}g_{tt}\big)}
 {g_{tt}\big(hI_1^2-{\cal{N}}^2(2\pi\alpha')^2\hat{g}_{22}g_{33}g_{tt}\big)}}\ I_1.
\ee %
According to AdS/CFT dictionary, it was shown in \cite{Karch} that
$I_1$ is a source term for $a_1$ at the boundary of AdS black hole.
More precisely the expectation value of $J_1$ on the boundary is
considered to be $I_1$. We assume that the standard AdS/CFT formulation
and hence the relation between $I_1$
and $J_1$ still exists in non-commutative background. $a_1(u)$ is
a real field and the only way for \eqref{A1} to remain real for all
values of $u$ is if both numerator and denominator change sign
at the same point \textit{i.e.} $u=u_*$. The reality condition of the
gauge field then yields to conductivity equations as %
\bse\label{NT}\begin{align} %
 &\label{NT1}g_{11}g_{tt}-(B^2+\hat{E}^2)=0,\\
 &\label{NT2}hI_1^2-{\cal{N}}^2(2\pi\alpha')^2\hat{g}_{22}g_{33}g_{tt}=0,
\end{align}\ese %
where all induced metric components are evaluated at $u_*$. At it is
seen from \eqref{NT1}, the $B$ field appears as an electric field
added to external field $E$. The backreaction of $B$ on the background
is already considered. One should check that the backreaction of $E$ is on the background
ignorable. This happens if the value of $E$ is smaller than the
value of $B$ on the brane \textit{i.e.} $E<B$ which is equivalent to
\be\label{condition} %
 \theta^3\hat{E}^2<1,
\ee %
for large values of $u$. The backreaction of $E$ is thus negligible if it satisfies \eqref{condition}.

After substituting induced metric components
and $B$ in \eqref{NT1}, one obtains %
\be %
 u^3_*(1-\theta^3u_*^3)-(1+\theta^3u_*^3)^2\hat{E}^2=0,
\ee %
whose roots are %
\be\label{root} %
 u^3_*=\frac{1-2\theta^3\hat{E}^2\pm\sqrt{1-8\theta^3\hat{E}^2}}{2\theta^3(1+\theta^3\hat{E}^2)}\ .
\ee %
where the reality of $u_*$ imposes
$\theta^3\hat{E}^2\leq\frac{1}{8}$ which is stronger than
\eqref{condition}. We use \eqref{NT2} and \eqref{root} to find
conductivity in terms of electric field which is finally given by
\footnote{For small value of $\theta$ one can extend \eqref{root}
yielding
\be %
u^3_*=\frac{1}{2\theta^3}\Big(1-2\theta^3\hat{E}^2\pm(1-4\theta^3\hat{E}^2)+O(\theta^6)\Big)\nonumber
\ee
Obviously minus sign reproduces commutative solution at zero temperature. }
\be %
 \sigma^2_{11}=\frac{I_1^2}{E^2}=(2\pi\alpha')^4{\cal{N}}^2\hat{E}^{4/3}(1+5\theta^3\hat{E}^2),
\ee %
up to $O(\theta^6)$. In this case non-commutativity increases the
value of conductivity. For the critical maximum possible electric
field E for a given $\theta$, $\theta^3\hat{E}^2=1/8$, the
conductivity is obtained as %
\be %
 \sigma^2_{11}=(2\pi\alpha')^4{\cal{N}}^2\frac{64\theta}{3^{5/3}}
\ee %
\subsection{High temperature case}
The high temperature non-commutative background is given by
\be\label{Nmetric2}\begin{split} %
ds^2=&
 u^{3/2}\bigg(h(u)\Big(f(u)dt_E^2+dx_1^2\Big)+dx_2^2+dx_3^2
 +dx_4^2\bigg)\cr
 +&u^{-3/2}\bigg(\frac{du^2}{f(u)}+u^2d\Omega_4^2\bigg),
 %h(u)&=\frac{1}{1+\theta^3u^3},\ \ \ \ \ e^{\phi}=g_su^{3/4}\sqrt{h(u)},\ \ \ \ \ B=B_{t1}=\theta^{3/2}u^3h(u)
\end{split}\ee %
where all other parameters are the same as \eqref{metric2} and \eqref{Nmetric1}.
In high temperature, as it was in pervious case, $A_1(u,t)$ is taken to be %
\be %
 A_1(t_E,u)=iEt_E+a_1(u).
\ee %
Hence the DBI action for $N_f$ D8-branes in the high temperature background is the same as \eqref{dbiaction}. It is
easy to see that conductivity equations also become the same as \eqref{NT}
%\be\begin{split} %
% S_{D8}=-\frac{8\pi^2N_fT_8}{3}\int dudt\frac{1}{\sqrt{h}}
% \sqrt{\hat{g}_{22}g_{33}\big[g_{tt}(\hat{a}_1'^2+g_{11}g_{uu})-g_{uu}(B^2+\hat{E}^2)\big]}\ .
%\end{split}\ee %
%The conserved charge associated with $A_1(u,t)$ is %
%\be\label{termalcharge} %
% I_1\equiv\frac{1}{\sqrt{h}}\frac{{\cal{N}}(2\pi\alpha')^2\hat{g}_{22}g_{tt}g_{33}a_1^{\prime}}
% {\sqrt{\hat{g}_{22}g_{33}\big[g_{tt}(\hat{a}_1'^2+g_{11}g_{uu})-g_{uu}(B^2+\hat{E}^2)\big]}}
%\ee %
%Solving \eqref{termalcharge} for the gauge field leads to%
%\be %
% \hat{a}_1^{\prime}=\pm\sqrt{\frac{hg_{uu}\big(B^2+\hat{E}^2-g_{11}g_{tt}\big)}
% {g_{tt}\big(hI_1^2-{\cal{N}}^2(2\pi\alpha')^2\hat{g}_{22}g_{33}g_{tt}\big)}}\ I_1
%\ee %
%Reality condition of the gauge field yields to conductivity equations as %
\bse\begin{align} %
 &\label{T1}g_{11}g_{tt}-(B^2+\hat{E}^2)=0\\
 &\label{T2}hI_1^2-{\cal{N}}^2(2\pi\alpha')^2\hat{g}_{22}g_{33}g_{tt}=0
\end{align}\ese %
After substituting induced metric components and $B$ \eqref{Nmetric2} in \eqref{T1}, one obtains %
\be %
 u^3_*(1-\theta^3u_*^3)-(1+\theta^3u_*^3)^2\hat{E}^2-u_T^3=0,
\ee %
whose roots are \footnote{Reality condition of $u_*$ imposes $\hat{E}^2\leq\frac{1-4\theta^3u_T^3}{4\theta^3(2+\theta^3u_T^3)}$.} %
\be\label{u*} %
 u^3_*=\frac{1-2\theta^3\hat{E}^2\pm\sqrt{1-8\theta^3\hat{E}^2-4\theta^3u_T^3-4\theta^6u_T^3\hat{E}^2}}{2\theta^3(1+\theta^3\hat{E}^2)},
\ee %
and by using \eqref{T2} conductivity becomes \footnote{For small
value of $\theta$, \eqref{u*} becomes
\be %
u^3_*=\frac{1}{2\theta^3}\Big(1-2\theta^3\hat{E}^2\pm(1-4\theta^3\hat{E}^2-2\theta^3u_T^2)+O(\theta^6)\Big)\nonumber
\ee where minus sign reproduces commutative solution at non-zero
temperature.}
\be\label{conductivity11} %
 \sigma^2_{11}=(2\pi\alpha')^4{\cal{N}}^2(\hat{E}^2+u_T^3)^{2/3}\bigg(1+\frac{\theta^3}{3E^2}(3\hat{E}+u_T^3)(5\hat{E}+3u_T^3)\bigg),
\ee %
up to $O(\theta^6)$. In the limit of commutative space
($\theta\rightarrow0$), conductivity is in perfect agreement with
\cite{Bergman} and it increases in presence of non-commutativity.
Two limits corresponding to \eqref{limit1} and \eqref{limit2} are now %
\begin{itemize}
  \item Zero external field \\
  \be %
   \sigma_{11}=(2\pi\alpha')^2{\cal{N}}u_T\sqrt{1+\frac{\theta^3u_T^6}{E}},
  \ee %
  In commutative case conductivity is proportional to $T^2$ instead of $T$ in \eqref{limit1}.
  Moreover in non-commutative space, correction term plays an important role in high
  temperature. Note that $\frac{\theta^3u_T^6}{E}$ must always be
  smaller than 1.
  \item Zero temperature  \\
  \be %
   \sigma_{11}=(2\pi\alpha')^2{\cal{N}}\hat{E}^{2/3}\sqrt{1+5(2\pi\alpha')^2\theta^3}.
  \ee %
  Comparing to $E^{1/2}$ in \eqref{limit2}, $E^{2/3}$ appears in conductivity equation in this case.
\end{itemize}
It is evident that all equations in this subsection reduce to pervious subsection equations by setting $u_T$ to zero.

\section{A more general non-commutative background}
Here we consider a more general non-commutative background given by %
\be\label{fullmetric}\begin{split} %
ds^2=&
 u^{3/2}\bigg(h(u)\Big(f(u)dt_E^2+dx_1^2\Big)+h^\prime(u)\Big(dx_2^2+dx_3^2\Big)
 +dx_4^2\bigg)\cr
 +&u^{-3/2}\bigg(\frac{du^2}{f(u)}+u^2d\Omega_4^2\bigg),\cr
 h(u)=&\frac{1}{1+\theta^3u^3},\ \ \ \ \ \ h^\prime(u)=\frac{1}{1+\theta^{\prime3}u^3},\ \ \ \ e^{\phi}=g_su^{3/4}\sqrt{h(u)h^{\prime}(u)}\cr
 B=&B_{t1}=\theta^{3/2}u^3h(u),\ \ \ \ \ B^\prime=B_{23}=\theta^{\prime 3/2}u^3h^\prime(u)
\end{split}\ee %
As before, a component of gauge field living on D8-branes is turned
on \footnote{One can introduce a magnetic field on the D8-brane as
\be\label{magnetic} %
 \mathbb{B}^i=\epsilon^{ijk}\Big(B_{jk}+(2\pi\alpha')F_{jk}\Big) %
\ee %
where in this case we have $E\parallel\mathbb{B}$ and
$B\parallel\mathbb{B}$.}
\be\label{gaugefield} %
 A_1(t_E,u)=iEt_E+a_1(u).
\ee %
The DBI action for $N_f$ D8-branes in background \eqref{fullmetric} becomes %
\be\label{fulldbiaction}\begin{split} %
 S_{D8}&=-{\cal{N}}\int dudt_E\frac{u^{1/4}}{\sqrt{hh'}}\cr
 &\times\sqrt{(g_{22}g_{33}+B'^2)\big[g_{tt}(\hat{a}_1'^2+g_{11}g_{uu})-g_{uu}(B^2+\hat{E}^2)\big]}\ ,
\end{split}\ee %
One can easily simplify the factor $g_{22}g_{33}+B'^2$ appearing in the action which leads to %
\be\label{factor} %
 h'(u)u^3,
\ee %
and by substituting \eqref{factor} in the DBI action \eqref{fulldbiaction}, we have %
\be\begin{split} %
 S_{D8}&=-{\cal{N}}\int dudt_E\frac{u^{1/4}}{\sqrt{h}}
 \sqrt{u^3\big[g_{tt}(\hat{a}_1'^2+g_{11}g_{uu})-g_{uu}(B^2+\hat{E}^2)\big]}\ ,
\end{split}\ee %
Note that $u^{1/4}\sqrt{u^3}$ is exactly the same as $\sqrt{\hat{g}_{22}g_{33}}$ in the
action \eqref{dbiaction} and it is consequently evident that the action \eqref{fulldbiaction} and
\eqref{dbiaction} are alike. This result shows that the effect of non-commutativity in "2" and "3"
directions can not be recognized by making ansatz \eqref{gaugefield}. In other words
 non-commutativity does not change the action and consequently the conductivity.
 The independence of the DBI action on the non-commutativity parameter has been also
 observed in \cite{Nakajima,Arean}.

We now try another ansatz which is \footnote{Due to the rotation
symmetry of the background in 23-plane, the other ansatz
$A_2(t_E,u)=iEt_E+a_2(u)$ will have the same result as
\eqref{gaugefield2}. Moreover by using \eqref{magnetic} we have
$E\perp\mathbb{B}$ and $B\parallel E$.}
\be\label{gaugefield2} %
 A_3(t_E,u)=iEt_E+a_3(u),
\ee %
and DBI action then becomes %
\be\label{fulldbiaction2}\begin{split} %
 S_{D8}&=-{\cal{N}}\int dudt_E\frac{u^{1/4}}{\sqrt{hh'}}\cr
 &\times\sqrt{g_{uu}\big[(g_{22}g_{33}+B'^2)(g_{tt}g_{11}+B^2)-g_{11}g_{22}\hat{E}^2\big]+g_{22}\hat{a}_3^{\prime2}(g_{tt}g_{11}+B^2)}\ .
\end{split}\ee %
The conserved charge associated to $a_3(u)$ is given by %
\be\label{I3}\begin{split} %
 I_3&\equiv\frac{1}{\sqrt{hh'}}\cr
 &\times\frac{\tilde{g}_{22}(g_{tt}g_{11}+B^2)\hat{a}_3^{\prime}}{
 \sqrt{g_{uu}\big[(g_{22}g_{33}+B'^2)(g_{tt}g_{11}+B^2)-g_{11}g_{22}\hat{E}^2\big]+g_{22}\hat{a}_3^{\prime2}(g_{tt}g_{11}+B^2)}},
\end{split}\ee %
where $\tilde{g}_{22}=u^{1/4}g_{22}$. From \eqref{I3}, it is easy to find $a_3(u)$ which is %
\be\label{A3} %
 \hat{a}_3^{\prime}=\pm\sqrt{\frac{hh'g_{uu}\big[(g_{22}g_{33}+B'^2)(g_{tt}g_{11}+B^2)-g_{11}g_{22}\hat{E}^2\big]}
 {(g_{tt}g_{11}+B^2)(g_{22}hh'I_3^2-(g_{tt}g_{11}+B^2)\tilde{g}_{22}^2)}}\ I_3.
\ee %
Conductivity equations coming from \eqref{A3} are %
\bse\begin{align} %
 &\label{FT1}(g_{22}g_{33}+B'^2)(g_{tt}g_{11}+B^2)-g_{11}g_{22}\hat{E}^2=0\\
 &\label{FT2}g_{22}hh'I_3^2-(g_{tt}g_{11}+B^2)\tilde{g}_{22}^2=0
\end{align}\ese %
\eqref{fullmetric} and \eqref{FT1} lead to %
\be %
 (1+\theta^3u_*^3)(u_*^3-\hat{E}^2)-u_T^3=0
\ee %
In the zero temperature case (\textit{i.e.} $u_T=0$)
non-commutativity, neither $\theta$ nor $\theta'$, does not affect
the conductivity and it is like the commutative case \cite{Bergman}.
But, in the thermal case we have %
\be\label{u*2} %
 u_*^3=\frac{-1+\theta^3\hat{E}^2\pm\sqrt{1+2\theta^3\hat{E}^2+\theta^6\hat{E}^4+4\theta^3u_T^3}}{2\theta^3},
\ee %
and conductivity becomes\footnote{ Similar to what was discussed in pervious sections, here only the plus sign is acceptable. }
\be\label{conductivity} %
 \sigma^2_{33}=(2\pi\alpha')^4{\cal{N}}^2(\hat{E}^2+u_T^3)^{2/3}(1-\frac{2}{3}\theta^3u_T^3)
\ee %
up to $O(\theta^6)$. \eqref{conductivity} indicates that $\theta'$
does not affect conductivity but $\theta$ decreases value of
conductivity. Comparing \eqref{conductivity11} and
\eqref{conductivity}, it is clearly seen that
non-commutativity parameter appears with opposite sign in these two
equations depending on which gauge field component is turned on.
Although in commutative background there is no difference among
$x_1,x_2$ and $x_3$ and therefore conductivity is the same for them
\textit{i.e.} $\sigma_{11}=\sigma_{22}=\sigma_{33}$, in
noncommutative background we have
$\sigma_{11}\neq\sigma_{33}(=\sigma_{22})$.

\section{Conclusion}
The effect of non-commutativity on conductivity was studied in this
paper. We considered low and high temperature non-commutative
Sakai-Sugimoto and $N_f$ D8-branes were then embedded in these
backgrounds. Various components of gauge fields living on D8-branes
lead to different values of conductivity in non-commutative
background. Our gauge field configurations do not recognize
non-commutativity in "2" and "3" directions but non-commutativity in
"1" and "2" directions changes the value of conductivity. Although
conductivities in different directions, \textit{i.e.} $x^i$, in
commutative background are the same, they are not alike in
non-commutative background.

\section{Acknowledgment}
We are grateful to thank M. M. Sheikh-Jabbari for useful discussions
and comments. We also like to thank K. Bitaghsir for helpful
discussion.

\end{document}